\documentclass[aps,prl,showkeys,reprint]{revtex4}

\usepackage[utf8]{luainputenc}
\usepackage{amsmath,braket,graphicx,siunitx}

\newcommand{\ee}{\ensuremath{\mathrm{e}}}
\newcommand{\ii}{\ensuremath{\mathrm{i}}}
\newcommand{\op}[1]{\ensuremath{\hat{#1}}}

\DeclareMathOperator{\sign}{sgn}

\begin{document}

\title{Unitary Two-State Quantum Operators Realized By Quadrupole Fields in the Electron Microscope}
\author{Stefan \surname{Löffler}}
\email{stefan.loeffler@tuwien.ac.at}
\affiliation{University Service Centre for Transmission Electron Microscopy, TU Wien, Wiedner Hauptstraße 8-10/E057-02, 1040 Wien, Austria}

\begin{abstract}
In this work, a novel method for using a set of electromagnetic quadrupole fields is presented to implement arbitrary unitary operators on a two-state quantum system of electrons. In addition to analytical derivations of the required quadrupole and beam settings which allow an easy direct implementation, numerical simulations of realistic scenarios show the feasibility of the proposed setup. This is expected to pave the way not only for new measurement schemes in electron microscopy and related fields but even one day for the implementation of quantum computing in the electron microscope.
\end{abstract}


\maketitle

\section{Introduction}

Unitary operators play a vital role across quantum mechanics and related fields as they model basis transformations. In transmission electron microscopy (TEM), the best-known such transformation is the Fourier transform which relates position space and reciprocal space and can be realized easily using a standard, round lens~\cite{WilliamsCarter1996}. Going from position space representation into reciprocal space representation allows the efficient determination of crystal structures and orientations with better accuracy and signal-to-noise ratio (SNR) than, e.g., when using high-resolution TEM images acquired in imaging mode. One primary reason for this is the fact that all electrons carrying a certain information --- e.g., about the lattice plane distance --- are focused in one spot in reciprocal space, while being distributed over the whole micrograph in position space. Thus, measuring a few electrons in a specific reciprocal space point already gives quantifiable information about the lattice plane spacing, whereas measuring the same (low) number of electrons in a position space image will just give a few counts scattered over the entire field of view.

Recently, an effective basis transformation was also employed to measure the orbital angular momentum (OAM) spectrum of an electron beam by means of a log-polar transformation \cite{NC_v8_i_p15536}. In that instance, too, a setup was found that transformed different OAM components in such a way that they showed up in unique measurement channels --- similar to diffraction spots ---, rather than producing small variations on an otherwise fairly large signal.

The idea of having a direct one-to-one correspondence between the intensity in a channel and the sought information is closely related to the concept of sparsity commonly found in compressed sensing applications (see, e.g., \cite{NL_v11_i_p4666,N_v11_i_p5617,IA_v6_p4875} and references therein) and blind source separation (see, e.g., \cite{Comon2010}). These methods, however, are post-processing techniques that in many cases require prior knowledge about the measured quantity. Above all else, however, their outcome strongly depends on the quality of the measured data, which in turn is heavily influenced by various noise sources, including shot noise and different electronic noise contributions in the read-out and processing components. Unitary operators, on the other hand, can be applied directly to a (quantum) system \emph{before} a measurement, thus allowing the measurement to be performed in a basis with optimal signal sparsity and SNR. The key requirement for this, however, is to find a way to perform the necessary unitary transformations directly in the instrument.

In this work, a setup is described that allows to realize arbitrary unitary operators on a two-state quantum system in a TEM. Two-state quantum systems are of particular importance as they model qubits, the building blocks of quantum computers. With the recent progress of the description of entanglement in the electron microscope \cite{U_v190_i_p39,Schattschneider2019}, the realization of arbitrary unitary operators on these qubits opens up many exciting possibilities for performing quantum computations in a TEM. In addition, the two-state system acts as an important model for the future development of setups for unitary operators on higher-dimensional systems.

\section{Theory}

Here, we use the vector space $\mathcal{V}$ spanned by the two orthonormal states $\ket{0}, \ket{1}$ given in position representation as
\begin{equation}
\begin{aligned}
\braket{\vec{r}|0} &= HG_{1,0}(\vec{r}) \propto x \cdot \ee^{-\frac{r^2}{w(z)^2}} \cdot \ee^{-\frac{\ii k r^2}{2R(z)}} \cdot \ee^{\ii \gamma(z)} \\
\braket{\vec{r}|1} &= HG_{0,1}(\vec{r}) \propto y \cdot \ee^{-\frac{r^2}{w(z)^2}} \cdot \ee^{-\frac{\ii k r^2}{2R(z)}} \cdot \ee^{\ii \gamma(z)},
\end{aligned}
\label{eq:HG}
\end{equation}
where $HG_{n,m}$ denotes the Hermite-Gaussian mode of order $(n, m)$ \cite{OC_v96_i_p123,U_v204_i_p27}, $w(z) = w_0 \sqrt{1+(z/z_R)^2}$ is the propagation-dependent beam size with the minimal beam waist $w_0 = \sqrt{2z_R/k}$ and the Rayleigh range $z_R$, $k$ is the wave number, $R(z) = z(1+(z_R/z)^2)$ is the curvature radius, and $\gamma(z) = \arctan(z/z_R)$ is the Gouy phase. Due to their primary orientation, $\ket{0}$ will be referred to as ``horizontal'' and $\ket{1}$ will be referred to as vertical in the following.

\begin{figure}
\includegraphics[width=8.6cm]{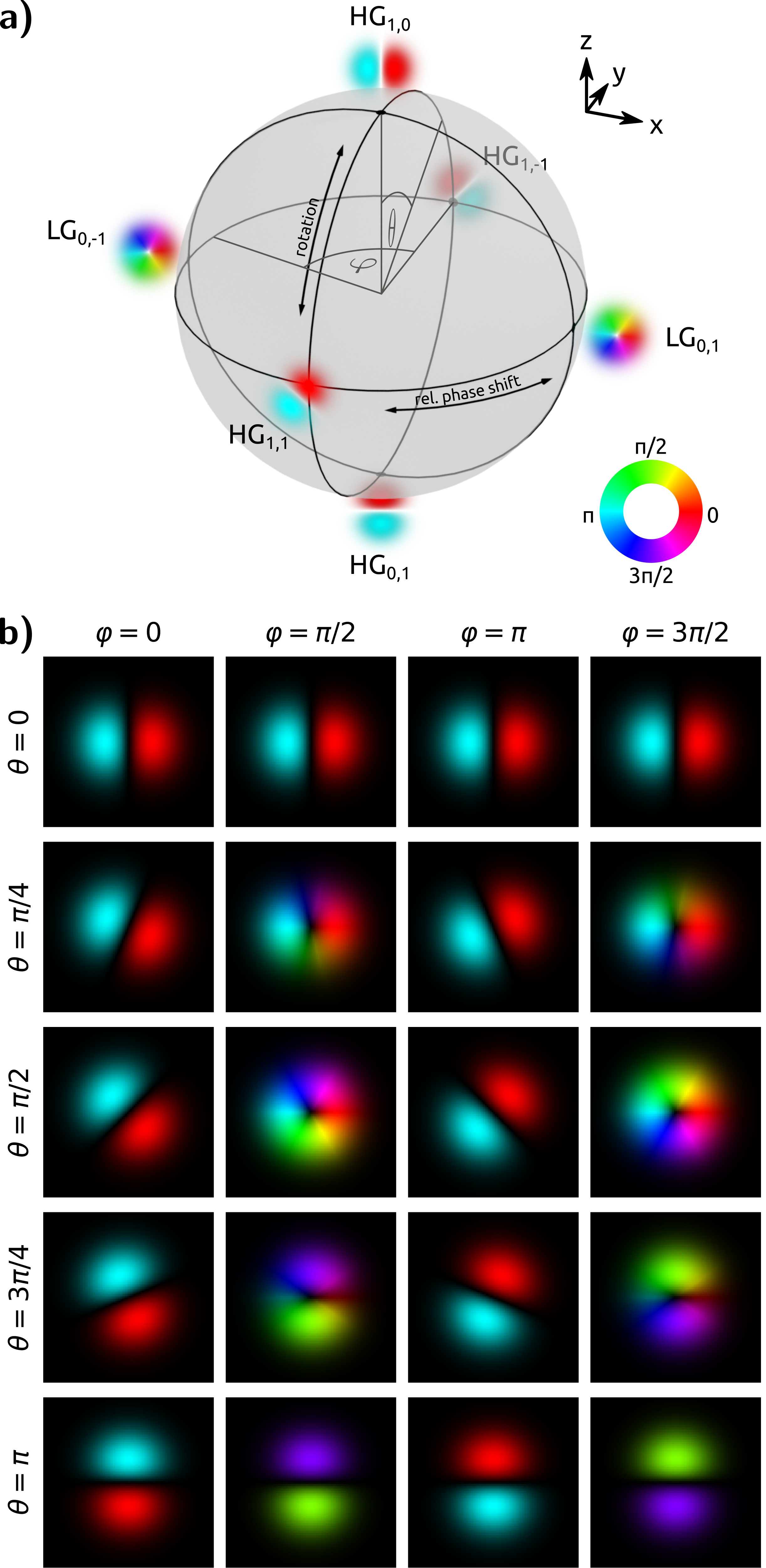}
\caption{a) Schematic of the Bloch sphere for the vector space described in the text. b) Selected states for various values of $\theta, \varphi$ according to eq.~\ref{eq:psi}. For all depicted states, intensity represents amplitude and color represents phase as indicated in the color wheel inset.}
\label{fig:bloch_sphere}
\end{figure}

Apart from a global phase factor, all normalized states $\ket{\psi} \in \mathcal{V}$ can be written as
\begin{equation}
	\ket{\psi} = \cos(\theta/2) \ket{0} + \sin(\theta/2) \ee^{\ii \varphi} \ket{1}
	\label{eq:psi}
\end{equation}
with $\theta \in [0, \pi], \varphi \in [0, 2\pi)$. Thus, all such states lie on the Bloch sphere (with the polar angle $\theta$ and the azimuthal angle $\varphi$) as depicted in fig.~\ref{fig:bloch_sphere}. Unitary operators are simply those changing $\theta$ and $\varphi$, i.e. rotations on the sphere. Following the scheme of (extrinsic) Euler angles, it is well-known that any arbitrary rotation can be decomposed into three successive rotations around cardinal axes, e.g. in the order $x$--$z$--$x$.

From fig.~\ref{fig:bloch_sphere} it can be seen that rotations around $x$ correspond to changing $\theta$. As is evident from eqs.~\ref{eq:HG} and \ref{eq:psi}, such an operation in the chosen basis corresponds to a rotation of the coordinate system in the plane perpendicular to the beam by an angle of $\theta/2$, i.e. $\vec{r} \mapsto \op{R}\vec{r}$, which can be realized in two ways: either one rotates the experimental setup (image, sample, etc.), which may even be achievable in post-processing in many cases, or one uses the well-known Larmor rotation \cite{JoMaMM_v324_i18_p2723,PRL_v110_i_p93601,NC_v5_i_p4586,U_v158_i_p17,PRX_v2_i_p41011} in the magnetic field of round lenses ubiquitous in electron microscopy.

The second ingredient to realizing arbitrary unitary operators on $\mathcal{V}$ is the ability to change $\varphi$, i.e., rotations around $z$ in fig.~\ref{fig:bloch_sphere}. From eq.~\ref{eq:psi}, it is evident that this corresponds to a relative phase shift between the two basis states. Here, a scheme for creating electron vortex beams (EVB) can be extended upon: the so-called ``mode conversion'' \cite{PRL_v109_i8_p84801,U_v204_i_p27}. Based on the idea of the optical mode converter \cite{OC_v96_i_p123}, it uses a set of two quadrupole lenses to convert a $HG_{1,1}$ beam into a $LG_{0,\pm 1}$ beam by means of a phase shift of $\delta\varphi = \pm \pi/2$.


\begin{figure}
\includegraphics[width=8.6cm]{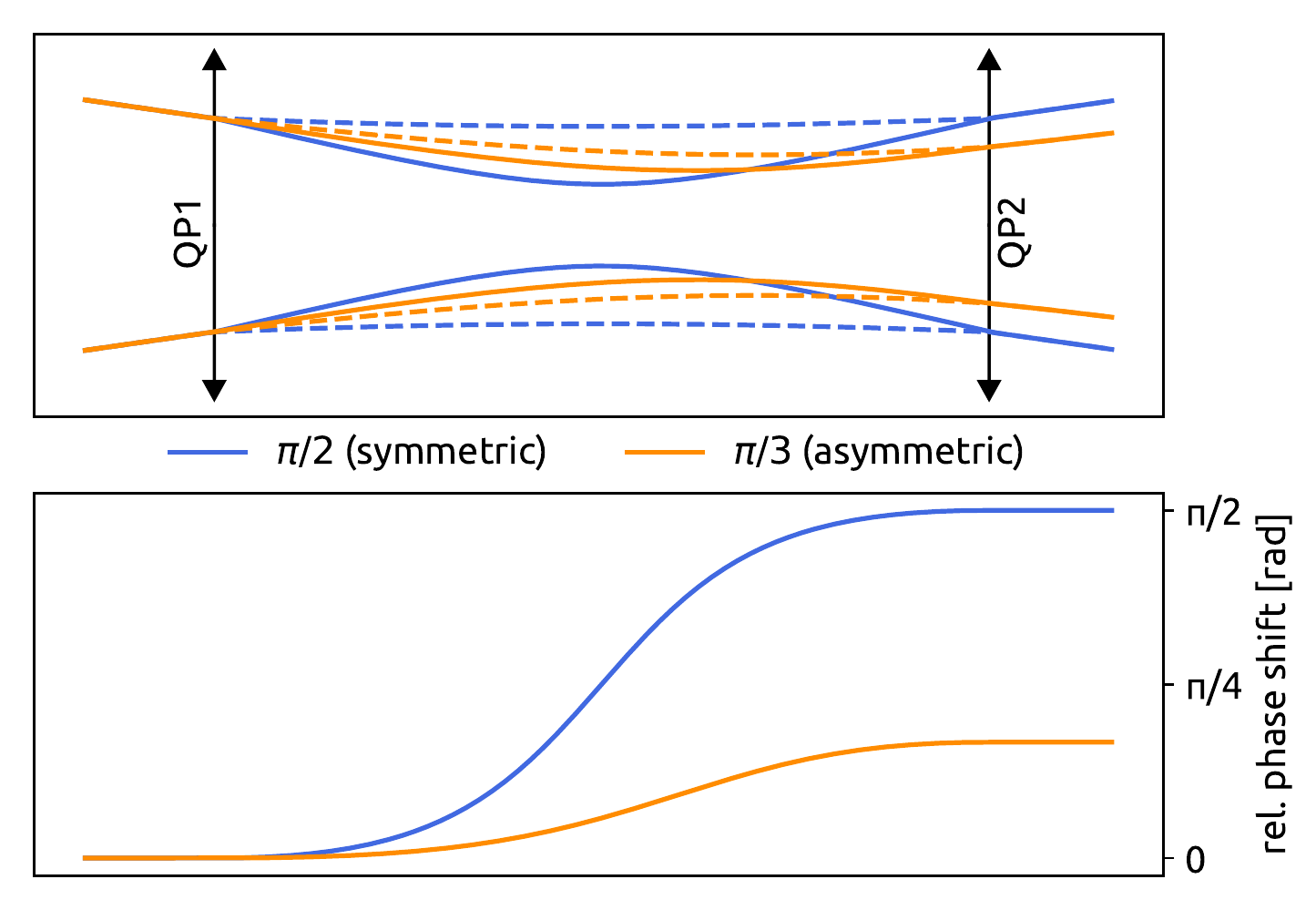}
\caption{Sketch of a relative phase shifter consisting of two quadrupoles (QP1, QP2). Two different settings leading to different phase shifts are shown (blue and orange). The top panel shows the horizontal (full lines) and vertical (dashed lines) beam diameters. The bottom panel shows the relative phase shift. In the shown scenario, the incident beam size was fixed.}
\label{fig:phase-shifter}
\end{figure}

Fig.~\ref{fig:phase-shifter} shows the principle setup of a relative phase shifter, consisting of two quadrupole lenses. Note that quadrupole fields can be rotated simply by changing the excitation of the four poles, thus making it easy to produce a relative rotation of $\theta/2$. The first quadrupole (QP1) produces an astigmatic beam from an incident round beam. The beam is focused in one direction (say, horizontally) before the second quadrupole (QP2), while it is defocused in the orthogonal direction. Due to this difference, the two basis states $\ket{0}, \ket{1}$ acquire different Gouy phase shifts, thus resulting in a relative phase shift by the time the two beams reach QP2. QP2 then has to be set up to compensate the action of QP1 and produce a non-astigmatic beam again.

To model the propagation of the beam through the QP lens setup, it is beneficial to introduce the complex beam parameter $q(z) = z - z_0 + \ii z_R$ for the two components, where $z_0$ is the position of the component's focus. Without loss of generality, $z_0 = 0$ will be assumed in the following. Using the complex beam parameter, all critical properties of the beam can be calculated:
\begin{equation}
w(z) = \sqrt{\frac{2|q|^2}{k \Im[q]}} \qquad R(z) = \frac{|q|^2}{\Re[q]} \qquad \gamma(z) = -\arg\left[\ii q\right]
\end{equation}
Additionally, both the propagation and the action of a lens can be modeled easily. Propagation over a distance $\delta z$ transforms $q \mapsto q + \delta z$, while a lens with focal length $f$ transforms $q \mapsto 1/(1/q - 1/f)$. A QP can then be modeled as a lens with focal length $f$ for one component and $-f$ for the other component.

The mode matching condition, i.e., the condition that the beam is round after QP2, results in the two conditions $w_h(z_2) = w_v(z_2)$ and $R_h(z_2) = R_v(z_2)$, where the subscripts $h, v$ denote the horizontal and vertical components, respectively, and $z_2$ is the position of QP2. The first of the two conditions ensures that the beam is round (non-astigmatic) at QP2, while the second condition ensures that it stays round even when propagating further after QP2. It is easily seen that mode matching is achieved if $q_h(z_2) = q_v(z_2)$ \cite{U_v204_i_p27}. A lengthy but straight-forward calculation shows that for two quadrupoles with focal lengths $f_1, f_2$ at a distance $d$, this can be achieved for an incident beam with
\begin{equation}
q_\text{in} = \frac{-df_1^2 + \ii d^2 f_1u}{f_1^2 + d^2u^2}
\quad\text{with}\quad
u = \sign[f_1]\sqrt{\frac{f_1f_2}{d^2} - 1},
\label{eq:mode-matching}
\end{equation}
with a relative phase shift of
\begin{equation}
\delta \varphi = -\arctan\left[\frac{2 u}{1 - u^2}\right].
\label{eq:phase-shift}
\end{equation}
Solving for $u$ gives
\begin{equation}
u = \frac{1 \pm \sqrt{1+\tan^2\delta\varphi}}{\tan\delta\varphi}.
\label{eq:u}
\end{equation}
This allows to calculate $u$ for any given relative phase shift $\delta\varphi$, where the sign has to be chosen appropriately for the quadrant (as given by eq.~\ref{eq:phase-shift}). This in turn fixes the relation between $f_1$ and $f_2$.

To sum up, for given $d$ and $\delta\varphi$, one ends up with three equations (one relating $u$ and $f_1, f_2$, and one each for the real and imaginary parts of $q_\text{in}$) for the four unknowns $f_1, f_2, \Re[q_\text{in}], \Im[q_\text{in}]$. Thus, one parameter can be chosen freely. In practice, this can be used to optimize the experimental conditions and to ensure, e.g., that the minimal and maximal excitation of the QPs are not exceeded and that the beam size is realizable with minimal aberrations.

Some properties of this setup are worth emphasizing: (i) in order to achieve mode matching, the incident beam must have a curvature at QP1 of $-d$, meaning that geometrically, it is focused at QP2; (ii) the phase shift implicitly depends on the size of the beam, i.e., while it is possible to choose either the incident or the outgoing beam size, it is not possible to choose both at the same time. In practice, both caveats can be worked around by the inclusion of a lens system before the quadrupole setup.

\section{Simulations \& Discussion}

\begin{figure}
\includegraphics[width=8.6cm]{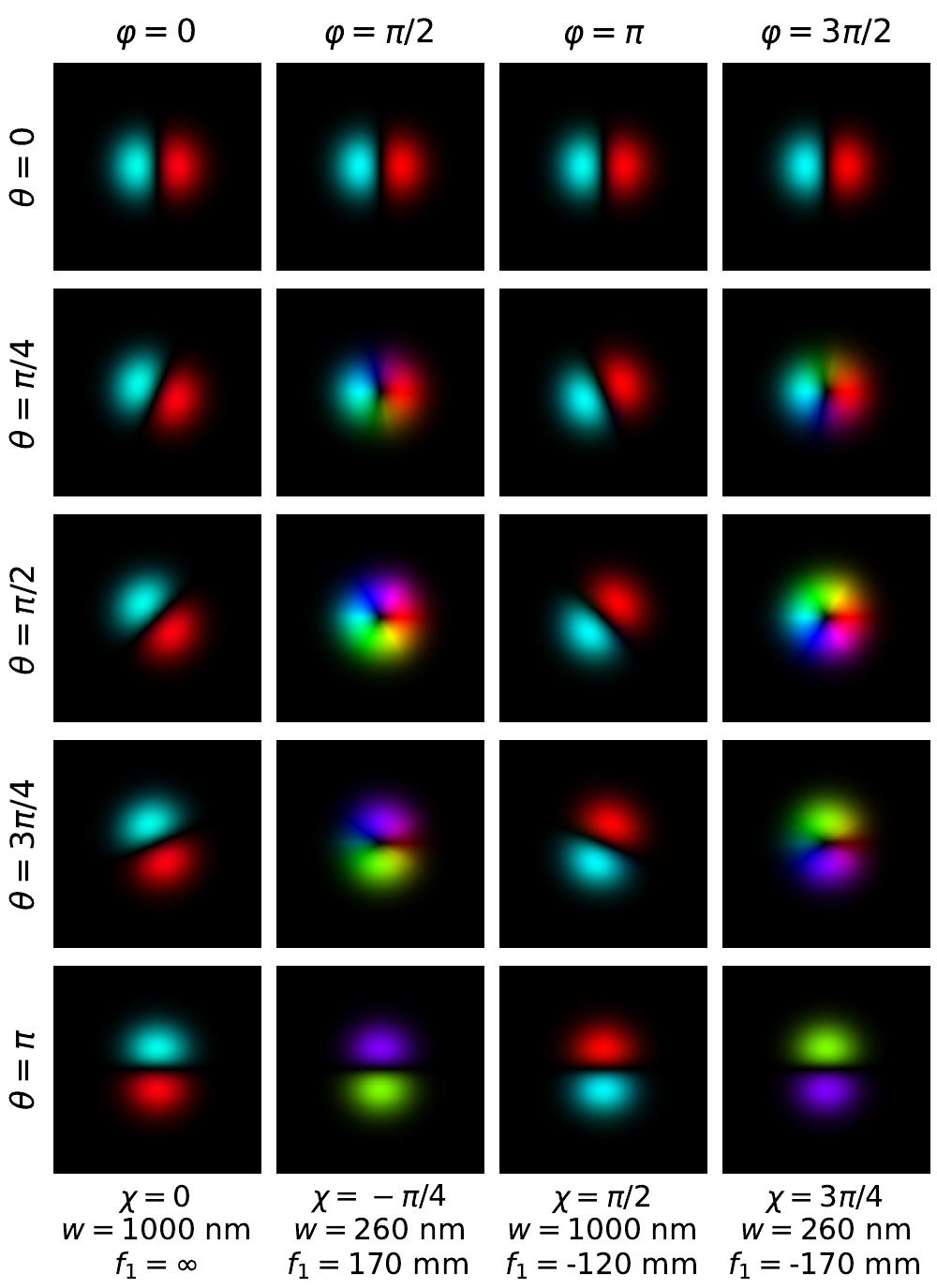}
\caption{Simulations of the phase shifter setup for the same values of $\theta, \varphi$ as in fig.~\ref{fig:bloch_sphere}. In all cases, the incident beam was a $HG_{1,0}$ beam rotated by $\theta/2$. For each $\varphi$, the incident beam size $w$, the QP1 focal length $f_1$, and the global phase compensation $\chi$ are indicated. For illustration purposes, a symmetric setup with $w_\text{in} = w_\text{out}$ and $f_1 = f_2$ was used. For $f_1 = \infty$, a numerical value of \SI{1}{\kilo\meter} was used. Amplitude and phase are shown as in fig.~\ref{fig:bloch_sphere}.}
\label{fig:sim}
\end{figure}

To corroborate the theoretical results, numerical simulations were performed using the \emph{virTUal TEM} software package \cite{U_v204_i_p27}. All simulations were performed for an incident $HG_{1,0}$ rotated by $\theta/2$ with an energy of \SI{200}{\kilo\electronvolt} using a setup as shown in fig.~\ref{fig:phase-shifter}. For simulation simplicity, an initially non-diffracting beam was transformed into a convergent beam using a round transfer lens before QP1. For clarity, a matching round lens after QP2 was included to flatten the phase front to ease comparability. The two QPs had a spacing of $d=\SI{120}{\milli\meter}$.

The results are summarized in fig.~\ref{fig:sim}. A comparison to fig.~\ref{fig:bloch_sphere} shows perfect agreement. It should be noted that in all cases except $\varphi=0$, the beam acquired a global phase $\chi$ as indicated in the figure. This stems from the propagation distance between the QPs, similar to the optical path length in light optics. As the global phase is inconsequential in this work (and can be compensated for by physical flight paths, lens systems, or temporarily changing the speed of the electrons), it is removed from the images in fig.~\ref{fig:sim} for better comparability.

Special care has to be taken for the edge cases $\varphi = 0$ and $\varphi = \pi$. For $\varphi = 0$, i.e. no phase shift, it suffices to switch off both QPs entirely. In the present simulations, the convergence angle of the incident beam was set to produce a (geometrical) focus at $d/2$. This resulted in a perfectly symmetric transfer of the incident plane to the exit plane. Since this only works in the geometrical limit of large beams, $w=\SI{1000}{\nano\meter}$ was chosen.

For $\varphi = \pi$, the prefactor of $\ket{1}$ has to be inverted, while $\ket{0}$ should be unaltered. This is achieved by tuning the QPs to a focal length of $-d$. Thereby, the beam goes through a horizontal line focus. Again, this works only in the geometrical limit, so $w=\SI{1000}{\nano\meter}$ was used.

For all other cases (i.e. $\varphi \notin \{0, \pi\}$), the phase shift was used to calculate $u$ (eq.~\ref{eq:u}), from which $f_1$ and $q_\text{in}$ were determined (eq.~\ref{eq:mode-matching}). It should be noted that it is possible to chain multiple phase shifters together, thus allowing to reduce even the fringe cases to ``normal'' situations, e.g. using $\varphi = 0 = \pi/2 -\pi/2$ and $\varphi = \pi = \pi/2 + \pi/2$.

In terms of practical applicability, the chosen parameters, while not specific to any particular instrument, are in a realistic order of magnitude range. Also the beam sizes of a few hundred nanometers are readily achievable in a TEM. As far as the incident beam is concerned, no perfect Gaussian beams have been produced to date, but sufficiently close approximations are possible \cite{PRL_v109_i8_p84801,NC_v5_i_p4586,U_v204_i_p27}.

\section{Conclusions}

In this work, a novel concept for using mode converters in the TEM was presented that allows the realization of arbitrary unitary operators on a two-state quantum system. This paves the way for the realization of higher-dimensional unitary operators, which in turn will open entirely new possibilities for electron microscopy and all fields it is applied in, from physics over material science and chemistry, to biology. Instead of post-processing data and looking for tiny signals in a huge, noisy background, the realization of unitary operators will allow much more efficient experiments by enabling scientists to devise measurement schemes where the electron beam is quantum-mechanically transformed into a basis in which the sought information can be read out directly. Moreover, together with the recent progress in understanding entanglement of free electrons, this work may well contribute one day to performing quantum computations in the electron microscope.

S.L. acknowledges fruitful discussions with Peter Schattschneider.

\bibliography{bibexport}

\end{document}